\documentclass[%
 reprint,
superscriptaddress,
 amsmath,amssymb,
 aps,
 prl,
floatfix,
]{revtex4-2}

\usepackage{graphicx}
\usepackage{dcolumn}
\usepackage{subcaption}
\usepackage{hyperref}
 \usepackage{url}
 \usepackage{ragged2e}
 \usepackage[strings]{underscore}
\usepackage{bm}
\usepackage{braket} 

\usepackage{xcolor}

\begin{document}

\preprint{APS/123-QED}

\title{Localization Contrast As a Fundamental Limit of Radiative Quantum Efficiency}

\author{Arsineh Apelian}
 \altaffiliation{aapelian@ucsb.edu}
\author{Tanay Tak}%
\affiliation{%
Materials Department, University of California, Santa Barbara, CA 93117 USA
}%
\author{Xuance Jiang }%
\affiliation{%
Department of Chemistry and Biochemistry, University of California, Santa Barbara, CA 93117 USA
}%
\author{Tsung-Yin Tsai }%
\affiliation{%
Materials Department, University of California, Santa Barbara, CA 93117 USA
}%
\author{James S. Speck}%
\affiliation{%
Materials Department, University of California, Santa Barbara, CA 93117 USA
}%
\author{Vojtech Vlcek}%
\affiliation{%
Materials Department, University of California, Santa Barbara, CA 93117 USA
}%
\affiliation{%
Department of Chemistry and Biochemistry, University of California, Santa Barbara, CA 93117 USA
}%



\date{\today}

\begin{abstract}
Disorder can localize valence- and conduction-band states to different degrees, yet whether such localization contrast has observable consequences remains poorly understood. Here, we show that localization contrast strongly impacts competing carrier-recombination channels. We illustrate this behavior in random In$_x$Ga$_{1-x}$N alloys. Increasing the In content from $x=13.5\%$ to $x=22\%$ progressively localizes the valence-band maximum state while leaving the conduction-band minimum state comparatively delocalized. This contrasting localization decreases the interband optical momentum matrix element underlying radiative recombination. For Auger-Meitner recombination, however, localization alone does not determine the coupling; atomic-scale disorder also breaks primitive-cell translational symmetry and modifies the momentum-space and phase structure of the states. This generates additional low-$Q$ components and enhances the Auger coupling by more than five orders of magnitude relative to pristine GaN. Thus, we provide a microscopic explanation for the competing trends in radiative and Auger recombination. More broadly, our results establish localization contrast as a mechanism through which disorder can control optical and carrier-carrier interactions in semiconductors. 

\end{abstract}

\maketitle



\textit{Introduction---} Disorder can localize electronic states, with wavefunction amplitudes that decay exponentially away from a region of spatial confinement, a phenomenon studied extensively in the context of electronic transport \cite{mott1979electronic,BERGMANN19841,PhysRevLett.56.980,Washburn01011986,AOKI19811079}. Importantly, disorder need not localize all electronic states to the same degree. In fact, considerable interest has focused on systems in which localized states coexist with more extended states, as in moiré materials, where their interplay remains an active area of study \cite{PhysRevB.85.195404, Yankowitz2012, 10.1021/acs.nanolett.4c03030, PhysRevLett.128.126401}. In these systems, the contrasting spatial distribution of the electronic states is engineered through the moiré superlattice \cite{Cao2020, Li2021, Regan2020, Wang2020}. Natively disordered alloys can likewise host localized and delocalized states in bulk \cite{PhysRevLett.25.520, PhysRevB.6.3598, PhysRevLett.86.2613, PhysRevB.88.125201}. Processes involving transitions between states with contrasting localization are therefore expected to exhibit new characteristic electronic and optical behavior because such processes can depend sensitively on the spatial structure of the wavefunctions.

In this work, we study such natively disordered systems that harbor a strong localization contrast at the band edges: InGaN alloys, which are widely used in light-emitting diodes (LEDs) and whose emission characteristics are strongly influenced by the interplay of alloying and disorder \cite{Humphreys2008, 10.1063/1.114359, CHICHIBU1999298, 10.1063/1.119013, Nakamura1995}. In these alloys, hole states at the valence-band edge are strongly localized, whereas electron states at the conduction-band edge remain comparatively delocalized \cite{10.1063/1.1405003,PhysRevApplied.20.044069,PhysRevB.91.035439,PhysRevApplied.20.064049}. At the same time, previous work has established the importance of radiative–Auger competition involving these band edge states for InGaN LED efficiency \cite{PhysRevB.95.125314, Delaney2009, PhysRevB.92.035207, 10.1063/1.4769374, Kioupakis_2013, Kioupakis2011, 10.1063/1.4965298, 10.1063/1.5002104, 10.1063/5.0006128, PhysRevB.105.195307, McMahon_2024, aufdermar}. Connecting these two effects at the microscopic level has remained challenging because explicit treatment of random-alloy disorder requires large supercells and sampling over multiple independent configurations. As a practical approach, previous studies have mostly focused on the special quasirandom structures (SQS) \cite{PhysRevLett.65.353} or the virtual-crystal approximation (VCA) \cite{PhysRevB.61.7877} to tackle the characteristic behavior of these alloys. 

As we show below, the intrinsic origin of the alloy optoelectronic behavior is closely linked to the microscopic detail of the individual electronic states involved in a given process. Here, we obtain single-particle wavefunctions of bulk InGaN alloys directly from first-principles and determine how this localization contrast affects radiative and Auger–Meitner recombination through matrix elements of the one-body momentum and two-body Coulomb operators. We show that the same atomic-scale disorder affects the two recombination processes very differently, producing a modest reduction in radiative recombination but a dramatic enhancement of Auger–Meitner recombination. This disparate response provides evidence that localization contrast can govern optoelectronic behavior and may therefore have broader implications for materials design in semiconductor systems. In InGaN, this mechanism contributes to the worsening efficiency of emitters at longer wavelengths, revealing a microscopic origin of the long-standing green-gap bottleneck \cite{Nakamura2009} in solid-state lighting.

\textit{Disorder-Induced Localization Contrast---} We first demonstrate how alloy disorder localizes the band-edge states of bulk In$_x$Ga$_{1-x}$N using first-principles calculations. We consider In compositions of $13.5\%$, $16\%$, and $22\%$, associated with violet, blue, and green wavelength emissions, respectively. Here, In atoms are randomly distributed over the cation sites, and pristine GaN ($0\%$ In) serves as the reference. For each composition, we use 15 random 864-atom supercells to capture configurational fluctuations in the electronic states.

We first quantify the contrasting localization using the participation ratio (PR), $ \mathrm{PR} = \frac{1}{\Omega}\, \frac{\left(\int_{\Omega} |\psi(\mathbf{r})|^{2}\, d^{3}r\right)^{2}} {\int_{\Omega} |\psi(\mathbf{r})|^{4}\, d^{3}r},$ which measures the spatial extent of the wavefunction within the supercell volume $\Omega$, with larger values indicating a more delocalized state. In GaN, both the valence band maximum (VBM) and conduction band minimum (CBM) are delocalized throughout the supercell. In the alloys, the VBM wavefunction exhibits strong localization in In-rich regions while the CBM remains comparatively delocalized, creating a localization contrast between the band-edge states. In both systems, the VBM exhibits N $2p$ orbital character and the CBM predominantly N $s$-like character. We attribute the stronger VBM localization primarily to fluctuations in the local chemical environment produced by the random In distribution. For a representative $13.5\%$ configuration, In substitution at the fixed pristine-GaN geometry reduces $\mathrm{PR}_{\mathrm{VBM}}$ by $\sim46\%$, while structural relaxation produces an additional $\sim13\%$ reduction relative to the unrelaxed alloy (see SI).

\begin{figure}[h]
    \centering
    \includegraphics[width=3.0in]{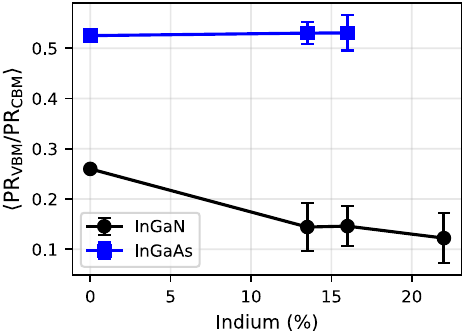}
    \caption{\justifying Configurationally averaged localization ratio $\left\langle \mathrm{PR}_{\mathrm{VBM}}/\mathrm{PR}_{\mathrm{CBM}}\right\rangle$ as a function of In composition for In$_x$Ga$_{1-x}$N (black) and In$_x$Ga$_{1-x}$As (blue). The ratio decreases with In content in InGaN, indicating that the valence edge localizes relative to the conduction edge but is constant in InGaAs over the range sampled. Error bars are standard deviations over 15 random configurations per composition for InGaN and 5 for InGaAs.}
    \label{Fig2}

\end{figure}
Figure~\ref{Fig2} shows the resulting configurationally averaged localization ratio $\langle \mathrm{PR}_{\mathrm{VBM}}/\mathrm{PR}_{\mathrm{CBM}}\rangle$. Our results demonstrate that alloy disorder reduces the PR of both band edges relative to pristine GaN, but acts far more strongly on the valence edge. At $13.5\%$ In, the $\mathrm{PR}_{\mathrm{VBM}}$ and $\mathrm{PR}_{\mathrm{CBM}}$ are reduced by $55\%$ and $19\%$, respectively, while at $22\%$ In the reductions increase to $63\%$ and $22\%$. The localization ratio (black curve) therefore decreases substantially with In content.

A natural question is whether this behavior is common to other widely used III–V alloys or is instead a distinctive feature of the nitrides. We choose InGaAs as a III–V semiconductor reference because it preserves the same In/Ga cation alloying while retaining a direct bandgap, providing a comparable system for analyzing valence- and conduction-band-edge localization. We therefore repeat the analysis for In$_x$Ga$_{1-x}$As at $x = 0\%, 13.5\%$ and $16\%$, with 5 random configurations for each alloy. As shown by the blue curve in Fig.~\ref{Fig2}, the localization ratio remains essentially constant as the In content increases, changing by only about $1\%$ relative to pristine GaAs. The localization effects that dominate the valence-band edge of InGaN are therefore absent in the arsenides alloy.

\textit{Localization and Radiative Recombination---}We now expose the consequences of this localization contrast for carrier recombination. Commonly, the internal quantum efficiency (IQE) of an LED is described within the ABC model as $\text{IQE} = B(n)n^2/[An+B(n)n^2+C(n)n^3]$, where $n$ is the carrier density and $A$, $B(n)$, and $C(n)$ are the effective Shockley--Read--Hall, radiative, and Auger-Meitner recombination coefficients, respectively \cite{ABC_IIIV,claude,David_2020}. Under charge-neutral carrier injection ($n\simeq p$), these terms reflect the number of carriers involved in each process: Shockley–Read–Hall recombination is effectively first order in carrier density, radiative recombination is second order, and Auger–Meitner recombination is third order. A reduction in IQE can arise from a decrease in the radiative coefficient $B(n)$ and/or an increase in the nonradiative coefficients $A$ and $C(n)$.

We first consider the radiative process, in which a conduction-band electron recombines with a valence-band hole and emits a photon. For these band-edge states, the low-energy quasiparticle problem can be represented by an approximately static single-particle Hamiltonian, whose eigenstates are orthogonal; this property is naturally preserved in our mean-field representation. We therefore examine how this localization contrast modifies the corresponding band-edge optical coupling. We characterize the radiative transition strength by the squared interband momentum matrix element,
\begin{equation}
S_{\mathrm{rad}}
\equiv
|\mathbf{p}_{vc}|^2
=
\sum_{\alpha=x,y,z}
\left|
\left\langle
\psi_v
\middle|
\hat{p}_{\alpha}
\middle|
\psi_c
\right\rangle
\right|^2 
\label{Srad}
\end{equation}
where $\hat p_\alpha$ is the $\alpha$-th component of the single-particle momentum operator. Within Fermi’s golden rule, $B_0\propto S_{\rm rad}$, so we use $S_{\rm rad}$ as a microscopic measure of the band-edge optical coupling, with the full rate additionally depending on energy and momentum conservation and carrier occupations. 

 \begin{figure}[h]
    \centering
    \includegraphics[width=3.0in]{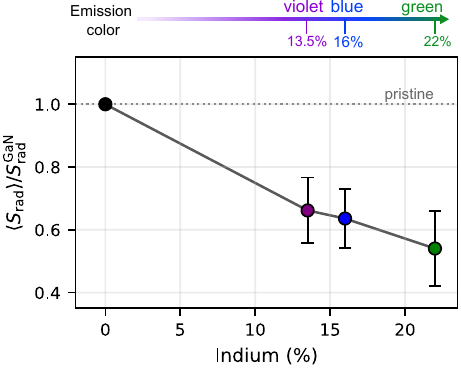}
    \caption{\justifying Configurationally averaged total radiative strength, $S_{\mathrm{rad}}=\sum_{\alpha=x,y,z}|p_{cv,\alpha}|^2$, normalized to its pristine GaN value, as a function of In composition. The band-edge radiative coupling decreases with increasing In content. Error bars are standard deviations over 15 random configurations per composition.}
    \label{Fig3}

\end{figure}

Both band edges lie at $\Gamma$ in these direct-gap alloys. We therefore compute $S_{\text{rad}}$ between the VBM and CBM states for the same 15 random configurations as the PR analysis. Figure~\ref{Fig3} shows the configurationally averaged $\left\langle S_{\mathrm{rad}}\right\rangle$, normalized to its pristine value, as a function of In. Our results show that the configurationally averaged radiative strength decreases monotonically with increasing In content, reaching a $46\%$ reduction at $x=22\%$ relative to pristine GaN. 

Both $\left\langle S_{\mathrm{rad}}\right\rangle$ and $\left\langle \mathrm{PR}_{\mathrm{VBM}}/\mathrm{PR}_{\mathrm{CBM}}\right\rangle$ therefore decrease with increasing In. For the $s$-like CBM and $p$-like VBM transition, the reduction in $\left\langle S_{\mathrm{rad}}\right\rangle$ arises because the momentum matrix element is determined by the spatial overlap of these states, and hence by their relative spatial extent, which is reduced as the hole wavefunction localizes while the electron wavefunction remains relatively delocalized. The radiative strength approximately follows $\langle S_{\rm rad}\rangle \approx K\langle \mathrm{PR}_{\rm VBM}/\mathrm{PR}_{\rm CBM}\rangle$, with $K=(0.814\pm0.013)(\hbar/a_0)^2 $ and 
$R^2=0.927$ across the three alloy compositions, demonstrating that the band-edge radiative transition follows the relative localization of the VBM and CBM single-particle states.

This establishes an intrinsic bulk-alloy reduction in the radiative strength. In practice, the active region of an In$_x$Ga$_{1-x}$N/GaN LED consists of quantum wells (QWs), in which a thin In$_x$Ga$_{1-x}$N alloy layer is confined between GaN barriers and grown along the $c$-axis of the wurtzite crystal \cite{nakamura2013blue, Akasaki1997, Akasaki_2006}. Built-in polarization fields along this direction spatially separate the electron and hole wavefunctions and reduce their envelope overlap \cite{CHICHIBU1999298,PhysRevB.57.R9435,10.1063/1.120269}. Therefore, the intrinsic bulk-alloy reduction identified here acts in addition to this effect, further suppressing the radiative coefficient $B(n)$.

\textit{Disorder and Auger-Meitner Recombination---} We next examine the Auger–Meitner process, in which the energy released by electron–hole recombination is transferred to a third carrier without accompanying photon emission. Unlike the one-body radiative transition, the Auger–Meitner process is a many-body transition involving four states coupled through the Coulomb interaction.

Within our mean-field treatment, the single-particle orbitals $\phi$ approximate the corresponding Dyson orbitals \cite{PhysRevA.50.4707, Romanova2022, HAMEL2002345}. We focus on the electron-hole-hole $(ehh)$ Auger process because the valence-edge manifold contains two nearly degenerate states, $\phi_v$ and $\phi_m$, needed to construct a set of single-particle states entering the direct and exchange Auger matrix elements. An analogous pair of near-degenerate conduction-edge states is not available for the electron-electron-hole ($eeh$) process. In the $ehh$ process, a conduction electron recombines with a valence hole and the released energy creates a hot hole deep in the valence band. We denote the CBM by $\phi_{c}$, the VBM by $\phi_{v}$, a second valence-edge state by $\phi_{m}$, and the final hot-hole state by $\phi_h$. The $ehh$ Auger process can be represented as the two-electron transition $\{\phi_h,\phi_c\}\rightarrow\{\phi_v,\phi_m\}$.

The Auger transition is constrained by both energy and momentum conservation. In our $\Gamma$-point supercell calculations, momentum conservation is treated implicitly within the supercell representation. Energy conservation places the final hot-hole state near $E_{\rm VBM}-E_g$. Using experimental bandgaps \cite{PhysRevB.95.144205}, we identify the calculated valence state $E_0$ nearest this target and include states satisfying $\Theta\!\left(\Delta E-|E_h-E_0|\right)=1$, where $\Theta(x)=1$ for $x\geq0$ and zero otherwise, with $\Delta E=20$ meV $\approx k_BT$ at room temperature. The microscopic Auger transition strength is given by \cite{PhysRevB.92.035207} 
\begin{equation}
S_{\mathrm{Auger}}
=
\sum_h
\left[
\left|W^{\mathrm{dir}}-W^{\mathrm{ex}}\right|^2
+
\left|W^{\mathrm{dir}}\right|^2
+
\left|W^{\mathrm{ex}}\right|^2
\right]
\end{equation} 
where the sum over $h$ is restricted to the selected states. For each allowed final hot-hole state $h$, the direct and exchange contributions account for the antisymmetry of the many-body wavefunction under fermion exchange and are defined as $W^{\mathrm{dir}} \equiv \langle \phi_v\phi_m|W|\phi_h\phi_c\rangle$ and $ W^{\mathrm{ex}} \equiv \langle \phi_v\phi_m|W|\phi_c\phi_h\rangle .$ We evaluate these matrix elements with either the bare interaction $v$, or with the screened interaction $W$ obtained within many-body perturbation theory at the random-phase approximation (RPA) level \cite{Romanova2023}. Since $C_0\propto S_{\rm Auger}$, we use $S_{\rm Auger}$ as a microscopic measure of the Auger strength; the full rate additionally depends on occupations and phase space.

 \begin{figure}[h]
    \centering
    \includegraphics[width=3.0in]{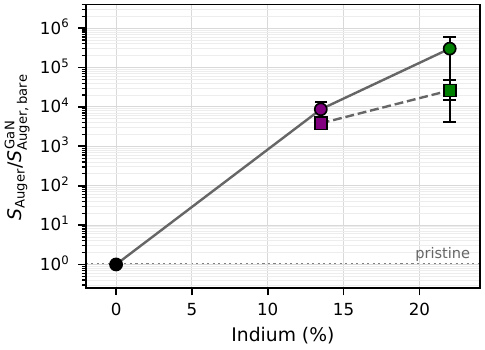}
    \caption{\justifying Configurationally averaged microscopic Auger strength, normalized to its pristine GaN value, as a function of In composition. Circles (solid line) and squares (dashed line) denote bare and screened results, respectively. At $22\%$ In, the bare Auger strength is enhanced by more than five orders of magnitude relative to pristine GaN, while the screened strength remains more than four orders above the same reference.
}
    \label{Fig5}

\end{figure}

Figure~\ref{Fig5} shows the configurationally averaged microscopic Auger transition strength, $\langle S_{\mathrm{Auger}}\rangle$, as a function of In content, for both bare and screened interactions. Both results are normalized to the same bare pristine GaN value, $S_{\mathrm{Auger,bare}}^{\mathrm{GaN}}$. Error bars indicate the sample standard deviation across the random configurations considered at each alloy composition; the details and stochastic uncertainties of the screened calculations are reported in the SI. Previous studies have linked disorder-induced relaxation of momentum-selection rules to enhanced Auger recombination \cite{PhysRevB.92.035207,PhysRevB.95.125314}. Here, we resolve this effect directly at the level of the microscopic Coulomb matrix elements: at $22\%$ In, the bare Auger transition strength is enhanced by more than five orders of magnitude relative to bare pristine GaN, while the screened strength remains more than four orders of magnitude above the same reference.

To resolve the microscopic origin of the enhancement, we analyze a representative exchange matrix element in reciprocal space, where $\phi_v$ and $\phi_c$ form the same band-edge pair as in the radiative transition. For simplicity, we focus only on the exchange contribution as its analysis is identical to the direct term. We use the bare Coulomb kernel $v$, since screening does not alter the resulting trend. For the recombining electron--hole pair, we define the orbital-pair amplitude as $\rho_{vc}(\mathbf{r}) = \phi_{v}^{*}(\mathbf{r})\,\phi_{c}(\mathbf{r})$ and its Fourier transform as $\rho_{vc}(\mathbf{Q}) = \int_{\Omega} d^{3}r\, \rho_{vc}(\mathbf{r})\, e^{-i\mathbf{Q}\cdot\mathbf{r}}$ where $\mathbf{Q}$ is the momentum transferred between the internal plane-wave components of the two orbitals. For the band-edge states, the low-energy static Hamiltonian gives mutually orthogonal states, yielding $\rho_{vc}(\mathbf{Q}{=}0) = \langle\phi_{v}|\phi_{c}\rangle = 0$. In the Auger channel, $\rho_{vc}$ enters through the Coulomb potential it generates, $\Phi_{vc}(\mathbf{r}) = \int_{\Omega} d^{3}r'\, v(\mathbf{r}-\mathbf{r}')\rho_{vc}(\mathbf{r}')$, which we decompose into its spatially uniform and fluctuating components,
\begin{equation}
\Phi_{vc}(\mathbf{r}) = \overline{\Phi}_{vc} + \delta\Phi_{vc}(\mathbf{r}).
\label{eq:Phidecomp}
\end{equation}
Due to state orthogonality, the spatially uniform component does not contribute to the transition: $V_{\rm exchange}=\int_{\Omega} d^3r\,\rho_{mh}(\mathbf r)[\overline{\Phi}_{vc}+\delta\Phi_{vc}(\mathbf r)]=\overline{\Phi}_{vc}\langle\phi_m|\phi_h\rangle+\int_{\Omega}d^3r\,\rho_{mh}(\mathbf r)\delta\Phi_{vc}(\mathbf r)=\int_{\Omega}d^3r\,\rho_{mh}(\mathbf r)\delta\Phi_{vc}(\mathbf r)$, where $\langle\phi_m|\phi_h\rangle=0$. Thus, $V_{\rm exchange}$ contains only the fluctuating, finite-$Q$ part of the Coulomb potential.

\begin{figure}
    \centering
    \includegraphics[width=0.95\columnwidth]{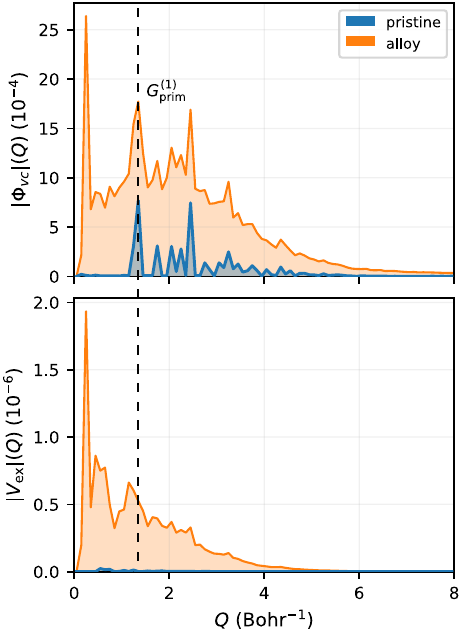}
\caption{\justifying
Reciprocal-space analysis of the Auger exchange contribution, with $Q\equiv|\mathbf Q|$. (top) Magnitude $|\Phi_{vc}|(Q)$ of the Coulomb potential generated by the VBM--CBM orbital-pair amplitude, where $\Phi_{vc}(\mathbf Q)=(4\pi/Q^2)\rho_{vc}(\mathbf Q)$. Alloy disorder introduces additional finite-$Q$ structure, including contributions below the first pristine reciprocal-lattice shell $G_{\mathrm{prim}}^{(1)}$, which are strongly weighted by the Coulomb kernel. (bottom) Magnitude of the exchange integrand, where $V_{\mathrm{ex}}(\mathbf Q) =\rho_{mh}(-\mathbf Q)\Phi_{vc}(\mathbf Q)$. These additional momentum-transfer channels contribute to the Auger exchange process underlying $V_{\mathrm{exchange}}$, however, contributions with different phases can partially cancel in the final sum over $\mathbf Q$.
}
    \label{fig:phi34_vexchange}
\end{figure}

In pristine GaN, primitive-cell translational symmetry gives $\rho_{vc}^{\mathrm{prim}}(\mathbf r+\mathbf R)=\rho_{vc}^{\mathrm{prim}}(\mathbf r)$, restricting $\rho_{vc}^{\mathrm{prim}}(\mathbf Q)$ to primitive reciprocal-lattice vectors $\mathbf G_{\mathrm{prim}}$. Since $\rho_{vc}^{\mathrm{prim}}(\mathbf Q=0)=0$, the first nonzero components occur at the first nonzero reciprocal-lattice shell (blue peaks in Fig. \ref{fig:phi34_vexchange}). For the alloy, breaking primitive-cell translational symmetry while preserving only the periodicity of the larger alloy supercell allows Fourier components at additional supercell momenta. The exact zero at $\mathbf{Q}=0$ is preserved by orthogonality, but substantial weight appears at small finite $Q$ below the first pristine reciprocal-lattice shell that are forbidden in pristine GaN, shown as orange peaks in the top panel of Fig.~\ref{fig:phi34_vexchange}. This behavior is nontrivial since disorder is typically associated with the loss of long-range order and a suppressed response near $Q\rightarrow0$. However, at any finite but small $Q$, the fluctuations can be present. These low-$Q$ components are particularly important because the Coulomb kernel weights them strongly through $4\pi/Q^2$, enhancing the effect of the potential fluctuations at small $Q$. 

The appearance of finite low-$Q$ components does not by itself determine the final matrix element (see bottom panel in Fig.~\ref{fig:phi34_vexchange}). The exchange contribution depends on the phase-dependent product $\rho_{mh}(-\mathbf Q)\delta\Phi_{vc}(\mathbf Q)$. Many of the large features in $\delta\Phi_{vc}(\mathbf Q)$ are reduced when multiplied by $\rho_{mh}(-\mathbf Q)$, and the sum over $\mathbf Q$ may exhibit partial cancellation. Still, the alloy retains a larger noncancelling contribution than pristine GaN. Moreover, the magnitude of the Auger matrix element is determined by the signed overlap of both orbital-pair amplitudes, explaining both the large pristine-to-alloy enhancement and its strong dependence on the random atomic configuration and selected deep-valence state.

\textit{Conclusion---}In summary, our results show that a strong localization contrast between electron and hole states can have a profound effect on the interactions that couple them: disorder suppresses the radiative matrix element while strongly enhancing the Auger matrix elements. With these findings, we demonstrate that the contrasting response of the momentum and Coulomb operators to disorder-modified wavefunctions constitutes an intrinsic mechanism governing the worsening efficiency of long-wavelength InGaN LEDs. In QWs, polarization fields and quantum confinement introduce additional effects beyond this bulk-alloy mechanism \cite{CHICHIBU1999298,PhysRevB.57.R9435,10.1063/1.120269}. This is particularly important as the QW envelope wavefunction overlap acts evenly on radiative and Auger processes \cite{10.1063/1.4769374}. This intrinsic contribution therefore explains the observed decrease of $B(n)$ relative to $C(n)$ at longer emission wavelengths \cite{PhysRevApplied.23.034002,schiavon,yichao2022}. More broadly, the compositional tunability of InGaN, which enables nitride emitters to span much of the visible spectrum \cite{Humphreys2008}, makes the efficiency reduction with increasing In content an \emph{inherent} limitation of this technologically important material system. Hence, long-standing attempts to circumvent the “green gap” problem \cite{Nakamura2009} must address a more fundamental limitation than a purely engineering obstacle. An important extension will be to determine how thermal and dynamic disorder further modify the localization contrast and the resulting matrix elements.

\textit{Acknowledgments---}
The authors thank Steven Crisostomo for fruitful discussions. This work was supported by a grant from the Alfred P. Sloan Foundation (FG-2024-22110). This work used Bridges-2 at Pittsburgh Supercomputing Center through allocation MAT260010, MAT260107, PHY250387 from the Advanced Cyberinfrastructure Coordination Ecosystem: Services \& Support (ACCESS) program, which is supported by National Science Foundation grants \#2138259, \#2138286, \#2138307, \#2137603, and \#2138296. Support for this work (JSS, TT, TYT) include the Simons Foundation, Grant No. 601952; the Vannevar Bush Faculty Fellowship managed by the Office of Naval Research, Award No. N00014-25-1-2040; and the Korea Institute for Advanced Technology(KIAT) through Seoul Viosys Co., Ltd. (Award No. SB250158).

\bibliography{main}

\end{document}